\documentclass[showpacs,preprintnumbers,amsmath,amssymb,nofootinbib,floatfix]{revtex4}

\usepackage[dvips]{graphicx}
\usepackage{dcolumn}
\usepackage{bm}

\def\mapgeq{\mathbin{\lower.3ex\hbox{$\buildrel>\over{\smash{\scriptstyle\sim}\vphantom{_x}}$}}}
\def\mapleq{\mathbin{\lower.3ex\hbox{$\buildrel<\over{\smash{\scriptstyle\sim}\vphantom{_x}}$}}}
\def\mapgeqeq{\mathbi{\lower.3ex\hbox{$\buildrel>\over{\smash{\scriptstyle\approx}\vphantom{_2}}$}}}
\def\mapleqeq{\mathbin{\lower.3ex\hbox{$\buildrel<\over{\smash{\scriptstyle\approx}\vphantom{_2}}$}}}
\mathchardef\hanaO="724F
\def\Journal#1#2#3#4{{#1} {\bf #2} (#4) #3}

\def\NPB{Nucl. Phys. B}

\def\NPSUPPL{Nucl. Phys. Proc. Suppl.}
\def\PLB{{Phys. Lett.} B}

\def\PLBOLD{Phys. Lett.}
\def\PRL{Phys. Rev. Lett.}
\def\RMP{Rev. Mod. Phys.}
\def\PRD{Phys. Rev. D}

\def\PTP{Prog. Theor. Phys.}
\def\JHEP{JHEP}

\def\EPJ{Euro. Phys. J. C}

\def\JETPUSSR{Sov. Phys. JETP}

\def\ZETP{Zh. Eksp. Teor. Fiz.}

\def\SCI{Science}
\def\APJ{Astrophys. J.}

\def\NJP{New J. Phys.}

\def\Erratum{Erratum-ibid}

\begin{document}


 \title{Flavor Neutrino Masses giving sin${\bf \theta_{13}}$=0}

\author{Keisuke Yuda}
\email{9asnm021@mail.tokai-u.jp}
\author{Masaki Yasu\`{e}}%
\email{yasue@keyaki.cc.u-tokai.ac.jp}
\affiliation{\vspace{3mm}%
\sl Department of Physics, Tokai University,\\
4-1-1 Kitakaname, Hiratsuka, Kanagawa 259-1292, Japan\\
}

\date{September, 2010}

\begin{abstract}
Among neutrino mixings, the reactor mixing angle, $\theta_{13}$, is observed to be almost vanishing and is 
consistent with $\theta_{13}=0$. 
We discuss how the condition of $\theta_{13}=0 $ constrains models of neutrino mixings
and show that, for  flavor neutrino masses given by $M_{ij}$ ($i,j$=$e,\mu,\tau$),
two conditions of $M_{e\tau } =  - e^{2i\gamma }\tan\theta_{23}M_{e\mu }$ and
$M_{\tau \tau } = e^{4i\gamma }M_{\mu \mu } + e^{2i\gamma }\left(2/\tan 2\theta_{23}\right)M_{\mu \tau }$
lead to $\theta_{13}$=0, where $\theta_{23}$ is the atmospheric neutrino mixing angle and 
$\gamma$ is its associated phase.  The rephasing invariance can select two phases provided by $\alpha$ = arg($M_{e\mu }$) and 
$\beta$ = arg($M_{e\tau }$), giving $\gamma=(\beta-\alpha)/2$. 
\end{abstract}

\maketitle
Flavor neutrinos, $\nu_{e,\mu,\tau}$, are mixed with each other during their flight, where neutrinos 
are described by mass-eigenstates, $\nu_{1,2,3}$ \cite{PMNS}. There are three distinct neutrino mixings, 
the atmospheric neutrino mixing \cite{SK,K2K}, the solar neutrino mixing \cite{OldSolor,Sun,Reactor} and the 
reactor neutrino mixing \cite{Theta13}, which are, respectively, denoted by three mixing angles, $\theta_{23}$, 
$\theta_{12}$ and $\theta_{13}$, for the $\nu_i$-$\nu_j$ mixing ($i,j$=$e,\mu,\tau$).
The masses $m_{1,2,3}$ and these mixing angles are currently constrained to be \cite{NuData}:
\begin{eqnarray}
\Delta m^2_\odot~\left[10^{-5}{\rm eV}^2\right] = 7.65
{\footnotesize
 {\begin{array}{*{20}c}
   { + 0.23}  \\
   { - 0.20}  \\
\end{array}}
},
\quad
\left|\Delta m^2_{atm}\right|~\left[10^{-3}{\rm eV}^2\right] = 2.40
{\footnotesize
{\begin{array}{*{20}c}
   { + 0.12}  \\
   { - 0.11}  \\
\end{array}}
},
\label{Eq:NuDataMass}
\end{eqnarray}
where $\Delta m^2_{atm}$, and $\Delta m^2_\odot$ are 
neutrino mass squared differences given by $\Delta m^2_\odot = m^2_2-m^2_1$ ($>0$ \cite{PositiveSolor}) for solar 
neutrinos, and $\Delta m^2_{atm} =  m^2_3-m^2_1$ for atmospheric neutrinos, and
\begin{eqnarray}
\sin ^2 \theta _{12}  = 0.304
{\footnotesize
{\begin{array}{*{20}c}
   { + 0.022}  \\
   { - 0.016}  \\
\end{array}}
},
\quad
\sin ^2 \theta _{23}  = 0.50
{\footnotesize
{\begin{array}{*{20}c}
   { + 0.07}  \\
   { - 0.06}  \\
\end{array}},
}
\quad
\sin ^2 \theta _{13}  = 0.01
{\footnotesize
 {\begin{array}{*{20}c}
   { + 0.016}  \\
   { - 0.011}  \\
\end{array}}.
}
\label{Eq:NuDataAngle}
\end{eqnarray}
The neutrino mixings are parameterized by the Pontecorvo-Maki-Nakagawa-Sakata (PMNS) unitary matrix 
$U_{PMNS}$ \cite{PMNS}, which converts the massive neutrinos $\nu_i$ ($i=1,2,3$) into 
the flavor neutrinos $\nu_f$ ($f=e,\mu,\tau$) : 
$\nu_f = \sum_{i=1}^{3}(U_{PMNS})_{fi}\nu_i$, and which is given by $U^{PDG}_{PMNS}=U^0_\nu K^0$ with
\begin{eqnarray}
U^0_\nu&=&\left( \begin{array}{ccc}
  c_{12}c_{13} &  s_{12}c_{13}&  s_{13}e^{-i\delta_{CP}}\\
  -c_{23}s_{12}-s_{23}c_{12}s_{13}e^{i\delta_{CP}}
                                 &  c_{23}c_{12}-s_{23}s_{12}s_{13}e^{i\delta_{CP}}
                                 &  s_{23}c_{13}\\
  s_{23}s_{12}-c_{23}c_{12}s_{13}e^{i\delta_{CP}}
                                 &  -s_{23}c_{12}-c_{23}s_{12}s_{13}e^{i\delta_{CP}}
                                 & c_{23}c_{13},\\
\end{array} \right),
\nonumber \\
K^0 &=& {\rm diag}(e^{i\phi_1}, e^{i\phi_2}, e^{i\phi_3}),
\label{Eq:Uu}
\end{eqnarray}
where $c_{ij}=\cos\theta_{ij}$ and $s_{ij}=\sin\theta_{ij}$ ($i,j$=1,2,3), as adopted by the 
Particle Data Group (PDG) \cite{PDG}. 
Leptonic CP violation is induced by one Dirac CP-violating phase ($\delta_{CP}$) and 
three Majorana phases ($\phi_{1,2,3}$) \cite{CPphases}, where
the Majorana CP-violating phases are determined by two combinations of $\phi_{1,2,3}$ such 
as $\phi_i-\phi_1$ ($i$=1,2,3). 

There are two distinct properties 
present in the observed data.  One is that the mixing 
angle $\theta_{13}$ is suppressed to show $\sin^2\theta_{13}\ll 1$ .  
The other is that 
$\Delta m^2_{atm}$ and $\Delta m^2_\odot$ show the hierarchy 
$\Delta m^2_\odot/\left|\Delta m^2_{atm}\right|\ll 1$.  Since the data are 
consistent with $\sin\theta_{13}=0$, a theoretical interest arises to find 
what conditions lead to $\sin\theta_{13}=0$ \cite{13-review}.  
There are known theoretical sources that lead to $\sin\theta_{13}=0$, 
which include the $\mu$-$\tau$ symmetry \cite{mu-tau}, the tri-bimaximal mixing scheme \cite{tribimaximal} and 
the strong scaling ansatz \cite{scaling}.  These examples call for specific relations among flavor neutrino masses.
However, as far as the condition of $\theta_{13}=0$ is concerned, they are over-constrained. 

In this note, we consider minimum requirement on flavor neutrino masses to yield $\sin\theta_{13}=0$.  
To do so, we use $U_{PMNS}$ with three Dirac phases, $\delta$, $\rho$ and $\gamma$, 
and three Majorana phases $\varphi_{1,2,3}$, instead of $U^{PDG}_{PMNS}$ to handle general 
phase structure of flavor neutrino mass matrix. Our $U_{PMNS}$ is parameterized by 
$U_\nu$ and $K$ \cite{BabaYasue} in place of $U^0_\nu$ and $K^0$:
\begin{eqnarray}
U_\nu&=&\left( {\begin{array}{*{20}c}
   1 & 0 & 0  \\
   0 & {e^{i\gamma } } & 0  \\
   0 & 0 & {e^{ - i\gamma } }  \\
\end{array}} \right)\left( {\begin{array}{*{20}c}
   {c_{12} c_{13} } & {s_{12} c_{13} e^{i\rho } } & {s_{13} e^{ - i\delta } }  \\
   { - c_{23} s_{12} e^{ - i\rho }  - s_{23} c_{12} s_{13} e^{i\delta } } & {c_{23} c_{12}  - s_{23} s_{12} s_{13} e^{i\left( {\delta  + \rho } \right)} } & {s_{23} c_{13} }  \\
   {s_{23} s_{12} e^{ - i\rho }  - c_{23} c_{12} s_{13} e^{i\delta } } & { - s_{23} c_{12}  - c_{23} s_{12} s_{13} e^{i\left( {\delta  + \rho } \right)} } & {c_{23} c_{13} }  \\
\end{array}} \right),
\nonumber \\
K &=& {\rm diag}(e^{i\varphi_1}, e^{i\varphi_2}, e^{i\varphi_3}).
\label{Eq:Uu_general}
\end{eqnarray}
The phases $\rho$ and $\gamma$ are redundant and can be removed by the phase redefinition.  
As a result, four phases in $U^{PDG}_{PMNS}$ are given by $\delta_{CP}=\delta+\rho$ and $\phi_1 = \varphi_1-\rho$ as well as 
$\phi_{2,3} = \varphi_{2,3}$.  
For the 2-3 rotation, one may choose the similar phase ($\tau$) to $\rho$ and $\delta$, 
contributing to $\delta_{CP}$ as $\delta_{CP} = \delta+\rho+\tau$. 
However, we have proved that $\gamma$ is a suitable phase for the 2-3 rotation \cite{BabaYasue}.
The phase $\tau$ can be removed by introducing a new definition: $\rho^\prime=\rho+\tau/2$,
$\gamma^\prime=\gamma+\tau/2$ and $\delta^\prime=\delta+\tau/2$
as well as $\varphi_2^\prime=\varphi_2-\tau/2$ and $\varphi_3^\prime=\varphi_3+\tau/2$.  As a result, 
we end up with the same definition of 
$\delta_{CP}$: $\delta_{CP}=\delta^\prime + \rho^\prime$. Therefore, the parameterization with 
$\delta^\prime$, $\rho^\prime$, and $\gamma^\prime$ gives 
Eq.(\ref{Eq:Uu_general}) as a general form of $U_{PMNS}$.

Let $M_\nu$ be a flavor neutrino mass matrix parameterized by
\begin{eqnarray}
M_\nu = \left( \begin{array}{*{20}{c}}
   a & b & c  \\
   b & d & e  \\
   c & e & f  \\
\end{array} \right),
\label{Eq:lmbda-x-y}
\end{eqnarray}
on the $(\nu_e, \nu_\mu, \nu_\tau)$-basis. Since $U_{PMNS}^TM_\nu U_{PMNS}={\rm diag}(m_1,m_2,m_3)$,
it is not difficult to derive
\begin{eqnarray}
&&
\left( e^{i\left( {\rho  - \delta } \right)}a - e^{i\left( {\rho  + \delta } \right)}\lambda _3 \right)\sin 2{\theta _{13}} + 2y\cos 2{\theta _{13}} = 0,
 \nonumber\\
&&
 \left( {{\lambda _1} - {\lambda _2}} \right)\sin 2{\theta _{12}} + 2x\cos 2{\theta _{12}} = 0,
\label{Eq:MixingAngles}\\
&&
 e\cos 2{\theta _{23}} - \frac{{{e^{ - 2i\gamma }}f - {e^{2i\gamma }}d}}{2}\sin 2{\theta _{23}} + {e^{ - i\left( {\rho  + \delta } \right)}}x\sin\theta_{13} = 0,
 \nonumber
\end{eqnarray}
as three vanishing off-diagonal elements \cite{BabaYasue-2}, where
\begin{eqnarray}
&&
{\lambda _1} = {e^{2i\rho }}\frac{{c_{13}^2a - s_{13}^2{e^{2i\delta }}{\lambda _3}}}{{c_{13}^2 - s_{13}^2}},
 \quad
 {\lambda _2} = c_{23}^2{e^{2i\gamma }}d + s_{23}^2{e^{ - 2i\gamma }}f - 2{s_{23}}{c_{23}}e,
 \nonumber\\
&&
 {\lambda _3} = s_{23}^2{e^{2i\gamma }}d + c_{23}^2{e^{ - 2i\gamma }}f + 2{s_{23}}{c_{23}}e,
\label{Eq:lmbda}\\
&&
 x = \frac{{{c_{23}}{e^{i\left( {\rho  + \gamma } \right)}}b - {s_{23}}{e^{i\left( {\rho  - \gamma } \right)}}c}}{{{c_{13}}}},
 \quad
 y = {s_{23}}{e^{i\left( {\rho  + \gamma } \right)}}b + {c_{23}}{e^{i\left( {\rho  - \gamma } \right)}}c.
\label{Eq:x-y}
\end{eqnarray}
If $\sin\theta_{13}=0$ is realized in Eq.(\ref{Eq:MixingAngles}), we obtain that
\begin{eqnarray}
&&
c =  - {e^{2i\gamma }}t_{23}b, 
\label{Eq:condition-b}\\
&&
f = {e^{4i\gamma }}d + e^{2i\gamma }\frac{{1 - t_{23}^2}}{{{t_{23}}}}e,
\label{Eq:condition-f}
\end{eqnarray}
where $t_{23}=\tan\theta_{23}$. The phase $\gamma$ turns out to be a phase difference between $b$ and $c$. 
We finally reach
\begin{eqnarray}
&&
M_\nu=
\left( {\begin{array}{*{20}{c}}
   a & {{e^{i\alpha }}\left| b \right|} & { - {e^{i\beta }}{t_{23}}\left| b \right|}  \\
   {{e^{i\alpha }}\left| b \right|} & d & e  \\
   { - {e^{i\beta }}{t_{23}}\left| b \right|} & e & {e^{2i\left( {\beta  - \alpha } \right)}}d + {e^{i\left( {\beta  - \alpha } \right)}}\frac{{1 - t_{23}^2}}{{{t_{23}}}}e  \\
\end{array}} \right),
\label{Eq:Mnu-13=0}
\end{eqnarray}
which provides the general structure of $M_\nu$ with $\sin\theta_{13}=0$ \cite{real-case}, where $\alpha$ and $\beta$, respectively, 
stand for phases of $b$ and $c$, determining 
\begin{eqnarray}
&&
\gamma  = \frac{\beta  - \alpha}{2}.
\label{Eq:gamma}
\end{eqnarray}
The rephasing invariance of $M_\nu$ allows us to set $a$, $d$ and $e$ to be real numbers.
Furthermore, $d$ is taken to be positive without loss of generality.  

One may wonder why Eqs.(\ref{Eq:condition-b}) and (\ref{Eq:condition-f}), which depend on the redundant phase $\gamma$, are appropriate.  
It is because $b$, $c$, $d$ and $f$ in $M_\nu$ is, respectively, transformed into 
$e^{i(\rho+\gamma)}b$, $e^{i(\rho-\gamma)}c$, $e^{2i\gamma}d$ and $e^{-2i\gamma}f$ after the redundant phases 
in $U_{PMNS}$ are removed.  Namely, for $U^{PDG}_{PMNS}$, the corresponding mass parameters get modified into
$b^{PDG}=e^{i(\rho+\gamma)}b$, $c^{PDG}=e^{i(\rho-\gamma)}c$, $d^{PDG}=e^{2i\gamma}d$, $e^{PDG}=e$ and 
$f^{PDG}=e^{-2i\gamma}f$ as obvious notations, which, give 
$y = {s_{23}}b^{PDG} + {c_{23}}c^{PDG}$ and 
$f^{PDG} = d^{PDG} + \frac{{1 - t_{23}^2}}{{{t_{23}}}}e^{PDG}$
without the apparent dependence of $\gamma$. See Ref.\cite{BabaYasue-2} for more details.
One may also wonder if $\tan2{\theta _{12}}$ given by Eq.(\ref{Eq:MixingAngles}) yielding
\begin{eqnarray}
\tan 2{\theta _{12}} &=&
\frac{{2{e^{i\left( {\rho  + \frac{{\alpha  + \beta }}{2}} \right)}}}}{{{c_{23}}}}\frac{{\left| b \right|}}{{{e^{i\left( {\beta  - \alpha } \right)}}\left| d \right| - {t_{23}}{\kappa _e}\left| e \right| - {e^{2i\rho }}{\kappa _a}\left| a \right|}},
\label{Eq:tan2-12}
\end{eqnarray}
remains real, where $\kappa_{a,e}$ take care of the sign of $a$ and $e$. Namely, phases must be cancelled out each other.
Since Eq.(\ref{Eq:tan2-12}) contains the phase $\rho$, we have to express $\rho$ in terms of the flavor neutrino masses.
It is convenient to use the relation  \cite{BabaYasue-2} given by
\begin{eqnarray}
&&
\rho = \arg (X),
\label{Eq:rho}
\end{eqnarray}
where $X$ is an analog of $x$ in Eq.(\ref{Eq:x-y}) but is determined by 
$U_{PMNS}^\dag M_\nu ^\dag {M_\nu }{U_{PMNS}}={\rm diag}(m^2_1,m^2_2,m^2_3)$. The parameter $X$ is
given by
\begin{eqnarray}
&&
X = \frac{{{c_{23}}{e^{i\gamma }}B - {s_{23}}{e^{ - i\gamma }}C}}{{{c_{13}}}},
\label{Eq:X}
\end{eqnarray}
where $B$ and $C$ are defined in
\begin{eqnarray}
&&
M_\nu ^\dag {M_\nu } = \left( {\begin{array}{*{20}{c}}
   A & B & C  \\
   {{B^ * }} & D & E  \\
   {{C^ * }} & {{E^ * }} & F  \\
\end{array}} \right).
\label{Eq:A-F}
\end{eqnarray}
Since $X$ is calculated to be
\begin{eqnarray}
&&
X = \frac{{{e^{i\frac{{\alpha  + \beta }}{2}}}\left| b \right|}}{{{c_{23}}}}\left[ {{\kappa _a}\left| a \right| + {e^{ - i\left( {\alpha  + \beta } \right)}}\left( {{e^{i\left( {\beta  - \alpha } \right)}}\left| d \right| - {t_{23}}{\kappa _e}\left| e \right|} \right)} \right],
\label{Eq:X-2}
\end{eqnarray}
one may replace ${{e^{i\left( {\beta  - \alpha } \right)}}\left| d \right| - {t_{23}}{\kappa _e}\left| e \right|}$ 
in Eq.(\ref{Eq:tan2-12}) by $X$ and we reach the following expression of $\tan 2{\theta _{12}}$:
\begin{eqnarray}
&&
\tan 2{\theta _{12}} = 
\frac{2}{{{c_{23}}}}\frac{{\left| b \right|}}{{\frac{{{c_{23}}{e^{ - i\rho }}X}}{{\left| b \right|}} - 2\cos \left( {\rho  - \frac{{\alpha  + \beta }}{2}} \right){\kappa _a}\left| a \right|}}.
\label{Eq:tan2-12-real}
\end{eqnarray}
Since $\rho=\arg(X)$, we prove that $\tan 2{\theta _{12}}$ certainly remains real and 
that our consideration using Eq.(\ref{Eq:Uu_general}) is correct.

The $\mu$-$\tau$ symmetric case, which provides $\theta_{13}=0$ and $\theta_{23}=\pi/4$, 
corresponds to $t_{23}=1$ in Eq.(\ref{Eq:Mnu-13=0}) leading to
\begin{eqnarray}
&&
M_\nu=
\left( {\begin{array}{*{20}{c}}
   a & {{e^{i\alpha }}\left| b \right|} & { - {e^{i\beta }}\left| b \right|}  \\
   {{e^{i\alpha }}\left| b \right|} & d & e  \\
   { - {e^{i\beta }}\left| b \right|} & e & {{e^{2i\left( {\beta  - \alpha } \right)}}d}  \\
\end{array}} \right),
\label{Eq:Mnu-13=0-mutau}
\end{eqnarray}
as well as $\alpha=\beta$. 
It should be noted that, if $\alpha\neq\beta$, the $\mu$-$\tau$ symmetry is broken but 
the prediction of $\theta_{13}=0$ and $\theta_{23}=\pi/4$ remains intact.  
This mass matrix is invariant 
under the interchange of ${\nu _\mu } \leftrightarrow  - {e^{i\left( {\beta  - \alpha } \right)}}{\nu _\tau }$.
This is the extended $\mu$-$\tau$ symmetry \cite{Koide}, which naturally manifests itself
in a special case of Eqs.(\ref{Eq:condition-b}) and (\ref{Eq:condition-f}).
The tri-bimaximal mixing scheme further predicts $\sin^2\theta_{12}$ to be 1/3 in Eq.(\ref{Eq:MixingAngles}).
The strong scaling ansatz is recovered by $e=-e^{i(\beta-\alpha)}t_{23}d$ leading to
\begin{eqnarray}
&&
M_\nu=
\left( {\begin{array}{*{20}{c}}
   a & {{e^{i\alpha }}\left| b \right|} & { - {e^{i\beta }}{t_{23}}\left| b \right|}  \\
   {{e^{i\alpha }}\left| b \right|} & d & -e^{i(\beta-\alpha)}t_{23}d  \\
   { - {e^{i\beta }}{t_{23}}\left| b \right|} & -e^{i(\beta-\alpha)}t_{23}d & e^{2i\left( {\beta  - \alpha } \right)}t^2_{23}d \\
\end{array}} \right).
\label{Eq:Mnu-13=0-SSA}
\end{eqnarray}
Another similar but new type of $M_\nu$ is given by $e=e^{i(\beta-\alpha)}d/t_{23}$ leading to
\begin{eqnarray}
&&
M_\nu=
\left( {\begin{array}{*{20}{c}}
   a & {{e^{i\alpha }}\left| b \right|} & { - {e^{i\beta }}{t_{23}}\left| b \right|}  \\
   {{e^{i\alpha }}\left| b \right|} & d & e^{i(\beta-\alpha)}d/t_{23}  \\
   { - {e^{i\beta }}{t_{23}}\left| b \right|} & e^{i(\beta-\alpha)}d/t_{23} & e^{2i\left( {\beta  - \alpha } \right)}d/t^2_{23} \\
\end{array}} \right).
\label{Eq:Mnu-13=0-another}
\end{eqnarray}

In summary, there are three important results found in our discussions:
\begin{itemize}
\item two conditions on $M_\nu$ to give $\sin\theta_{13}=0$ consisting of
\begin{eqnarray}
&&
M_{e\tau} =  - {e^{2i\gamma }}t_{23}M_{e\mu}, 
\quad
M_{\tau\tau} = {e^{4i\gamma }}M_{\mu\mu} + e^{2i\gamma }\frac{{1 - t_{23}^2}}{{{t_{23}}}}M_{\mu\tau},
\label{Eq:condition-c-f}
\end{eqnarray}
where $M_{ij}$ ($i,j$=$e,\mu,\tau$) is an $i$-$j$ matrix element of $M_\nu$,
\item the ``natural" emergence of the extended $\mu$-$\tau$ symmetry that 
arises from the requirement of 
$M_{e\tau} =  - {e^{2i\gamma }}M_{e\mu}$ and $M_{\tau\tau} = {e^{4i\gamma }}M_{\mu\mu}$ as in Eq.(\ref{Eq:Mnu-13=0-mutau}),
where the phase $\gamma$ breaks the exact $\mu$-$\tau$ symmetry, and
\item Eq.(\ref{Eq:Mnu-13=0-another}) as a new type of $M_\nu$ derived by $M_{\mu\tau}=e^{2i\gamma}M_{\mu\mu}/t_{23}$, 
which is similar to $M_\nu$ with $M_{\mu\tau}=-e^{2i\gamma}t_{23}M_{\mu\mu}$ for the strong scaling ansatz.
\end{itemize}
Since the appearance of the Dirac CP-violation requires $\sin\theta_{13}\neq 0$, 
we may add the following terms to $M_\nu$ to break the relations Eqs.(\ref{Eq:condition-b}) and (\ref{Eq:condition-f}):
\begin{eqnarray}
&&\Delta M_\nu
=
\left( {\begin{array}{*{20}{c}}
   0 & 0 & {\delta c}  \\
   0 & 0 & 0  \\
   {\delta c} & 0 & {\delta f}  \\
\end{array}} \right). 
\label{Eq:Mnu-13=0-breaking}
\end{eqnarray}
These presumably small parameters yield a nonvanishing value of $y$, thereby, giving $s_{13}\neq 0$ 
as follows:
\begin{eqnarray}
&&
\delta c = {e^{ - i\left( {\rho  - \gamma } \right)}}\frac{y}{{{c_{23}}}},
\quad
\delta f = {e^{ - i\left( {\delta  + \rho  - 2\gamma } \right)}}\frac{{2{s_{13}}x}}{{\sin 2{\theta _{23}}}}.
\label{Eq:Mnu-13=0-delta}
\end{eqnarray}
We will consider phenomenological analysis based on $M_\nu+\Delta M_\nu$
 and see effects of the leptonic CP violation in our future publication \cite{Future}.

\vspace{3mm}
\noindent
\centerline{\small \bf ACKNOWLEGMENTS}

The authors are grateful to T. Kitabayashi for useful discussions.


\end{document}